\def\t{\thinspace}
\def\msol{{\rm\thinspace M${}_\odot$}}
\def\fig{{\rm\thinspace Figure}}
\def\kms{{km s${}^{-1}$}}
\def\etal{{\it et al.\thinspace}}
\def\eg{{\it e.g.\ }}

\def\ie{{\it i.e.\ }}

\def\msolyr{{\rm\thinspace M${}_\odot\; {\rm yr}^{-1}$}}
\def\gcm3{{g cm${}^{-3}$}}

\def\msol{\hbox{$\rm\thinspace M_{\odot}$}}
\def\msolyr{\hbox{$\rm\thinspace M_{\odot} \; {\rm yr}^{-1}$}}

\def\h50{\hbox{$\rm\thinspace h_{50}$}}
\def\h50m1{\hbox{$\rm\thinspace h_{50}^{-1}$}}

\def\msol{\hbox{$\rm\thinspace M_{\odot}$}}
\def\etal{{\it et al.\thinspace}}
\def\eg{{\it e.g.\ }}

\def\ie{{\it i.e.\ }}

\def\fig{figure}

\def\p3m{P${}^3$M}
\def\ap3m{AP${}^3$M}
\def\-{{\em{---}}}

\documentstyle[emulateapj,psfig,apjfonts]{article}

\received{}
\accepted{}
\journalid{}{}
\articleid{}{}

\lefthead{Thacker \& Couchman}
\righthead{Supernovae Feedback}

\begin{document}

\title{Star Formation, Supernovae Feedback and the Angular Momentum
Problem in Numerical CDM Cosmogony: Half Way There?}

\author{R. J. Thacker\altaffilmark{1} and H. M. P. 
Couchman\altaffilmark{2}}

\altaffiltext{1}{Department of Astronomy,
University of California at Berkeley,
Berkeley, CA, 94720.}
\altaffiltext{2}{Department of Physics and Astronomy,
McMaster University, 1280 Main St. West, Hamilton, Ontario, L8S 4M1,
Canada.}

\begin{abstract}
We present a smoothed particle hydrodynamic (SPH) simulation that
reproduces a galaxy that is a
moderate facsimile of those observed. The primary failing point of
previous simulations of disk formation, namely excessive transport of
angular momentum from gas to dark matter, is ameliorated by the inclusion
of a supernova feedback algorithm that allows energy to persist in the
model ISM for a period corresponding to the lifetime of stellar
associations.  The inclusion of feedback leads to a disk at a redshift
$z=0.52$, with a specific angular momentum content within 10\% of the
value required to fit observations. An exponential fit to the
disk baryon surface density gives a scale length within 17\% of the
theoretical value. Runs without feedback, with or without star 
formation, exhibit the drastic angular momentum 
transport observed elsewhere.
\end{abstract}



\section{Introduction} 
A fundamental assumption of the analytic work on disk formation of Fall \&
Efstathiou (1980) is that during protogalatic collapse the baryonic
component conserves its specific angular momentum (AM). Pioneering N-body
hydrodynamic simulations (\cite{NB91}) showed that baryon cores, which
develop naturally within CDM simulations, systematically lose specific AM
to dark matter halos during merger events, contrary to the assumption of
Fall \& Efstathiou.  As detailed in White (1994), the resolution of this
perceived `numerical problem' is widely believed to be the inclusion of
feedback from supernovae (although for an alternative view see
\cite{MF99}), which is necessary in CDM cosmologies to avoid the cooling
catastrophe (\cite{WF91}). To date, simulations incorporating
feedback\-using a variety of algorithms to model the process\-have been
largely unsuccessful in preserving the specific AM of the baryons (\eg
\cite{nk,NW93,NS00}). Models, in which the gas is artificially prevented
from cooling before a `cooling epoch' (\cite{WE98}) have been considerably
more successful. Alternatively, Dominguez-Tenreiro \etal (1998), argue
that the inclusion of star formation can help stabilize disks against bar
formation, and subsequent AM loss. Sommer-Larsen \& Dolgov (2000) have
shown that warm dark matter (WDM)  does not suffer as significant an AM
deficit problem due to the reduction in short-scale power compared with
CDM.

This {\em Letter} presents results of simulations which model star
formation and feedback using an algorithm tested in detail in Thacker \&
Couchman (2000\-TC00). We achieve significant success in reproducing
fundamental properties of observed galaxies: the model galaxy does not
suffer a catastrophic ($\sim$80\%) loss of AM relative to the dark matter and
has a disk scale length comparable with those observed and in accord with
theoretically predicted values (\eg \cite{MO98}).

\begin{figure*}
\vspace{90mm}
\includegraphics{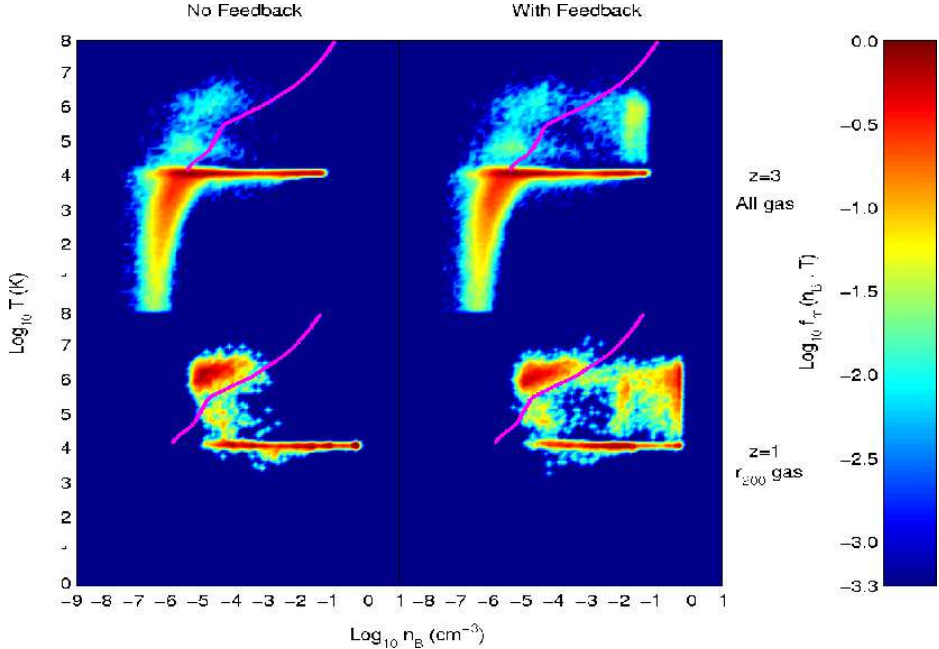}
\caption{Density--temperature filling functions for the NF
run (left) and ES (right)  at z=3 (top) and z=1 (bottom). The magenta
lines separate gas which can cool (below the line) and cannot cool
(above the line) by $z=0$ at each epoch ($t_{cool}=t_0-t$).} 
\label{fig1}
\end{figure*}

\section{Numerical Limitations of Feedback Models} 
It is not currently possible to model individual star formation and
feedback events in cosmological simulations of disk formation. The
result of limited numerical resolution is a large disparity between the
minimum simulation timescale and the true physical timescale
associated with supernova feedback. Physically, following a supernova,
the gas cooling time very rapidly increases to $O(10^7)$ years over
the time it takes the shock front to propagate. Thus it is the
timescale\-$t_{\dot E}$\-for the cooling time to increase markedly
which is of critical importance. In the SPH simulation, ``stars'' form
and ``supernova feedback'' occurs in gas cores which are typically
less than $h_{min}/5$ in diameter ($h_{min}$ being the effective
spatial resolution). Any rearrangement of the particles in such a
small region leads to a density estimate varying by at most $\sim7$\%.
Since the cooling time in the model is dependent on the local SPH
density, any energy deposited into a feedback region only reduces
the local cooling rate by increasing the temperature; the SPH
density cannot respond on a timescale which in real supernova events
would very rapidly increase the cooling time. A successful feedback
algorithm in an SPH model must thus overcome the fact that
$\rho_{SPH}$ does not change on the same time-scale as $t_{\dot E}$
and consequently that cooling times for dense gas cores at
$T<10^{6.5}$ K and $\rho > 2\, n_B$ cm${}^{-3}$ remain short ($<10$
Myr, as emphasized in \cite{SL98}).

\section{Algorithm and Initial conditions}
We model a standard Cold Dark Matter model: $\Omega_b=0.1$,
$\Omega_{CDM}=0.9$, $h=0.5$, $\sigma_8=0.6$ and shape parameter
$\Gamma=0.41$ with the adiabatic Bond \& Efstathiou (1984) CDM power
spectrum. The same candidate halo as discussed in Thacker (1999) was
selected for resimulation from a $100^3$ dark-matter-only simulation of
comoving width 48 Mpc. The halo, of mass $1.66\times 10^{12}$ \msol\, lies
on a filament $\sim 2$ Mpc long, and does not have a violent merger
history at low resolution.

Three SPH simulations were run, one with no star formation or feedback
(NSF), one with star formation but no feedback (NF) and one with both star
formation and feedback (ES: Energy Smoothing, see below).  The initial
conditions for each simulation were prepared by creating four mass
hierarchies in radial shells within the $48^3\,$Mpc$^3$ (comoving)
simulation volume.  The central high resolution region is 6 Mpc in
comoving diameter and has a total mass of $7.8\times 10^{12}$ \msol\ and a
particle number of $2\times$65,454 yielding particle masses of $1.2\times
10^7$ \msol\, and $1.1\times 10^8$ \msol\, for gas and dark matter
respectively. The minimum `glob' and dark halo mass resolutions are 52
times higher, corresponding to the number of SPH neighbors. The dark
matter particle mass is small enough to avoid spurious two-body heating of
the gas particles (\cite{SW97}). Only the high resolution region includes
SPH particles, which were given an initial temperature of 100 K. The
particle positions were drawn from a `glass' and the power spectrum was
truncated at the Nyquist frequency of each hierarchy.  A Plummer softening
length of $\epsilon=3$ kpc was chosen and the minimum SPH smoothing length
was $h_{min}=3.52$ kpc. 

The star formation and feedback algorithms are described in detail in
TC00, the key features are summarized here. Radiative cooling is included
for a gas of metallicity 0.01 $Z_\odot$. The star formation rate is
calculated using a Lagrangian Schmidt Law: $\dot{M}_* = \sqrt{4\pi
G\rho_{g}} \; c^*M_{g} = c^*M_g /t_{freefall}$. The SFR normalization (or
star formation efficiency) was set to $c^*=0.06$, 50\% higher than the
local group estimate of Gnedin (2000). Each gas particle carries an
associated `star mass' and can spawn two star particles each of half the
mass of the original gas particle.  This procedure leads to a delay
between star formation and the associated feedback (see TC00 for a full
discussion).  After spawning the first star particle, the gas particle
mass is decremented accordingly. The Energy Smoothing algorithm of TC00
(ESa) smoothes $5\times 10^{15}$ erg g${}^{-1}$ of feedback energy over
the SPH neighbor particles after a star formation event and allows this
energy to persist in an adiabatic state for a time $t_{1/2}=30$ Myr, after
which the energy is allowed to radiate away. The 30 Myr period is
motivated to coincide with the lifetime of stellar associations
(\cite{G97}), and is longer than the 5 Myr value used in TC00. Star
formation is only allowed in regions where the baryons are partially self
gravitating, $\rho_b>0.2\,\rho_{DM}$, and the local flow is converging
$(\nabla.{\bf v}<0)$. 

It is worth noting that since the adiabatic period in the ES algorithm,
$t_{1/2}$, is not a function of resolution, the (resolution dependent)
Courant condition does not guarantee a sufficiently short time-step for
feedback regions to expand. We have found that $dt<0.1 t_{1/2}$ is
necessary. This criterion imposes a maximum time-step length of 2 Myr
although in practice the time-step is limited by other criteria. At lower
resolution this criterion would become significant, which may partially
explain the comparatively poor results for the ES feedback algorithm
observed in TC00, where the algorithm was observed to have only a minor
impact on AM values.

\section{Results} 
The simulations were integrated to $z=0.52$, which required
15,930(15,900;17440) timesteps for the ES(NF;NSF) runs respectively,
requiring between 9 and 14 days on a 4 node Alphaserver ES40. Minimum
time-step values were $dt_{min}=$0.12 Myr for all simulations, and
average values were $dt=$0.44(0.44;0.40) Myr. At $z=0.52$, the virial
radius was $r_{200}=203.4\pm{0.4}$ kpc, and particles from the second
mass hierarchy had reached $1.5\,{r_{200}}$ preventing reliable
further integration.  The relaxation argument in Thomas \& Couchman
(1992) implies that 4 particles per softening volume is necessary to
avoid two-body heating in the dark matter at $z=0.52$, which is
satisfied in the densest regions of the simulation. At this epoch,
there were 14,546(14,466;14,403), 8,350(5,805;12,713) and
13,193(17,264;0) dark matter, gas and star particles within
$r_{200}$. At this SPH resolution shock processes should be
represented with moderate accuracy within the halo
(\cite{SM93}). Halo, disk and bulge masses are summarized in
Table~1. Disk edges are found using a surface density cut as in TC00,
with a slightly lower limit of $7\times10^{12}$ \msol Mpc${}^{-2}$ due
to the increased mass resolution.  At $z=0.52$, the disk edges defined
in this way were at 29(19;13)\t kpc. A comparison of the gas content
shows that at this epoch the ES galaxy has 3.1 times more gas than the
NF run.

\bigskip
\begin{tabular}{lcccc}
\hline\hline
\vspace{1mm}
Run & $M_h$ & $M_b$ & $M_d$ & B:D \\
 &   ($10^{12}$ \msol) & ($10^{10}$ \msol) & ($10^{10}$ \msol) & \\
\hline \vspace{1mm}
{\Large ${}^{ }$} ES & $1.56$ & $6.69$ & $3.94$ & 1.69:1\\
{\Large ${}^{ }$} NF & $1.55$ & $6.54$ & $3.26$ & 2.01:1\\
{\Large ${}^{ }$} NSF & $1.55$ & $7.98$ & $2.15$ & 3.71:1\\

\hline
\smallskip
\end{tabular}
{\small {\sc Table}~1.---Halo, disk and bulge masses, and bulge:disk 
ratios for the simulations.}
\medskip

Although the first gas cores are in place by $z=9$, star formation begins
at $z=5.6$ and the first feedback events occur at $z=3.95$.  A peak SFR of
51 \msolyr\ is reached for the NF run at $z=2.75$, while for the ES run
the peak is 27 \msolyr\ at $z=3.42$.  Due to feedback the ES SFR is
suppressed by up to 35 \msolyr\ compared to the NF run between $z=3.4$ and
$z=1.6$. At $z=0.52$ the average SFRs over the previous 0.4 Gyr are 1.67
\msolyr\ for the ES run and 0.96 \msolyr\ for the NF run. By $z=1$ in the
ES run 98\% of the bulge mass is in place and formation of the extended
gas disk begins largely after this epoch. The morphology of the ES galaxy
is type S0, \ie the system has a distinct disk and bulge component while
the NF galaxy is closer to E7, with a thin gaseous disk embedded within
it. 

{\epsscale{0.95}
\plotone{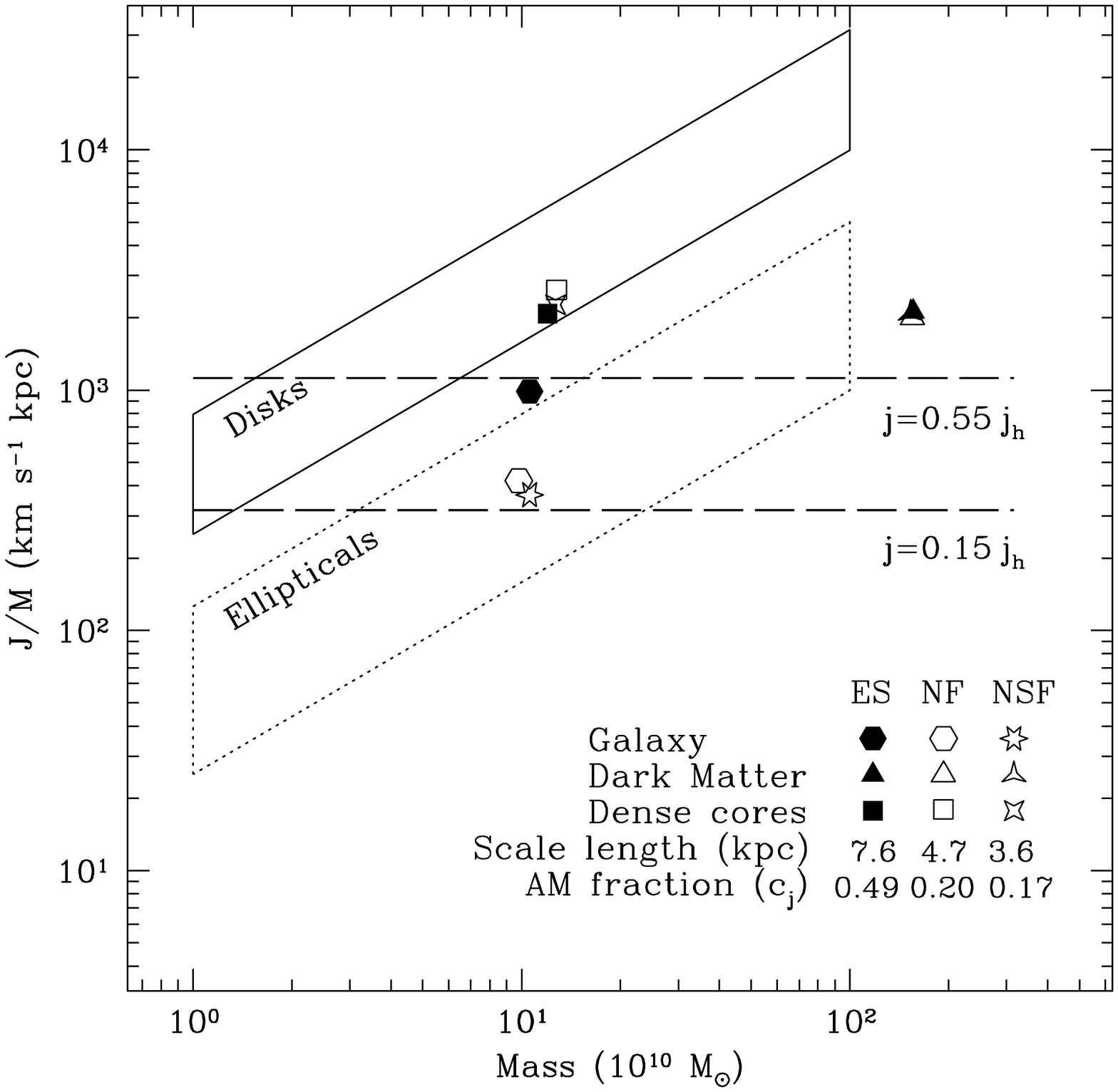}}

{\small {\sc Fig.}~2.---Specific AM for various components
of the simulations at z=0.52. The ES simulation achieves a $j$
value close to the combined observational and theoretical constraint $j_d
\simeq 0.82\,j_h$. Boxes are inferred from data assembled by Fall (1983).
Scale lengths are derived from the exponential fits to the disks
described in the text and $c_j=j_d/j_h$ values are also given.}

\subsection{Effect of Feedback on Overcooling} 
In \fig~\ref{fig1} we plot the mass filling factor for the gas $f_m(n,T)$,
defined as the fraction of mass per unit logarithmic interval at density
$n$ and temperature $T$. The integrated, constant density, cooling curves
are also shown to separate gas which can cool from that which cannot. The
z=3 plot for the NF run shows hot, low-density (halo)  gas ($T\sim 10^6$
K, $n_B\sim 10^{-5}$ cm${}^{-3}$), cooled gas cores ($T\sim 10^4$ K, $n_B
> 10^{-3}$ cm${}^{-3}$), and cold void gas ($T<10^4$ K, $n_B < 10^{-6}$
cm${}^{-3}$). The ES simulation develops a new hot, high-density phase,
which at $z=3$ has $T\simeq10^6$ K, $n_B\simeq 0.01$ cm${}^{-3}$, and
$\log f_m \simeq -1.3$ indicating that only a small amount (about the same
as the hot halo) of the gas is in this phase. The `bridge' ($f_m > -2.0$)
between the hot-high-density phase and the hot-low-density phase (to the
left of the cooling line) shows that some of the gas is being heated
sufficiently and rising high enough in the halos to have $t_{cool}>5$ Gyr.
This highlights the point that gas does not have to be blown-away to make
it unavailable for disk formation. However, it is clear from the $z=1$
plot that the predominant evolutionary pathway for feedback is recycling
of gas back into the cold, high-density phase for which cooling times
remain less than 1 Gyr. 

\subsection{Angular momentum and disk scale length}
Navarro \& Steinmetz (2000) argue, following work by Mo \etal (1998) and
using the observations of Mathewson \etal (1992) and Courteau (1998), that
the ratio of specific AM between disk and halo components, $c_j$, should
be given by,
\begin{equation}
c_j={j_d  \over j_h} \simeq {0.023 \over \lambda} \left( {\Delta \over
200} \right) 
\left( {V_c \over V_{200}} \right)^2
\end{equation}
at $z=0$, where $\Delta$ is the average overdensity within the virial
radius, $V_c$ is the disk circular speed, $V_{200}$ is the circular speed
at the virial radius, and $\lambda$ is the spin parameter of the halo.
Thus, if a disk galaxy within a halo of overdensity 200 has the same $V_c$
as $V_{200}$ and $\lambda=0.05$, then the disk component preserves one
half of its initial specific AM. For the ES simulation, $V_{200}=191$
\kms, $\lambda=0.075$, and $V_c=258$ \kms\ at $r=40$ kpc which,
extrapolated to $z=0$, predicts $c_j=0.55$. Parameters for the NF run are
identical (to within 2\%).

In \fig~2 we plot the specific specific AM of the halo, galaxy
(bulge+disk) and dense cores at $z=0.52$ for both simulations.  The most
significant result is that the specific AM of the entire ES galaxy is only
10\% lower than the predicted value of $c_j=0.55$. Both the NF and NSF
runs have minimal amounts of AM. This does not agree with the findings of
Dominguez-Tenreiro \etal (1998), who argue that star formation can help
prevent the formation of an angular-momentum-robbing bar. However, since
the NSF simulation does not form a bar, our argument is not conclusive. 
As expected, the bulge components (not plotted)  exhibit an extreme
deficit of specific AM ($j_{bulge}\sim 0.01 j_h$)  having been formed
primarily from first generation infall that has been subjected to little
feedback (even in the ES run).  The dense cores, defined as all the
baryons within $r_{200}$ satisfying $\delta_{b}>2000$ (\ie the galaxy plus
satellites), have a high specific AM value because they include the
contribution of one large core at 0.8$r_{200}$ (160 kpc)  that heavily
biases the ${\bf r}\times {\bf v}$ calculation. A large fraction of this
AM will be shed as the core merges with the central galaxy. The presence
of this satellite highlights the argument of Binney, Ortwin \& Silk
(2001): if one can expel the low angular momentum core, then baryons
falling in from large radii, which naturally have a large specific AM,
will produce a galaxy with a large specific AM value. 

Mo \etal (1998) have constructed an analytic model of disk formation that
assumes a negligible exponential disk embedded within an isothermal
dark halo. They parameterize the fraction of specific AM retained within
the baryons in terms of $c_j$ and find a disk
scale length given by, 
\begin{equation} 
R_d={ \lambda G M^{3/2}_h \over 2 V_{200} |E_h|^{1/2} } \; c_j. 
\end{equation} 
For our simulated halos $|E_h|=3.45\times 10^{52} J$
which implies $R_d=13.2\,c_j$ kpc. Using the $c_j$ values in 
\fig~2, the ES disk has an
expected scale length of $\sim 6.5$ kpc, while for the NF and NSF runs the
value is $\sim 2.5$ kpc. Least-squares exponential fits to our disks
(measured from $\epsilon$ to the disk edge) give $R_d=$7.6(4.7;3.6)\t kpc.

{\epsscale{0.95}
\plotone{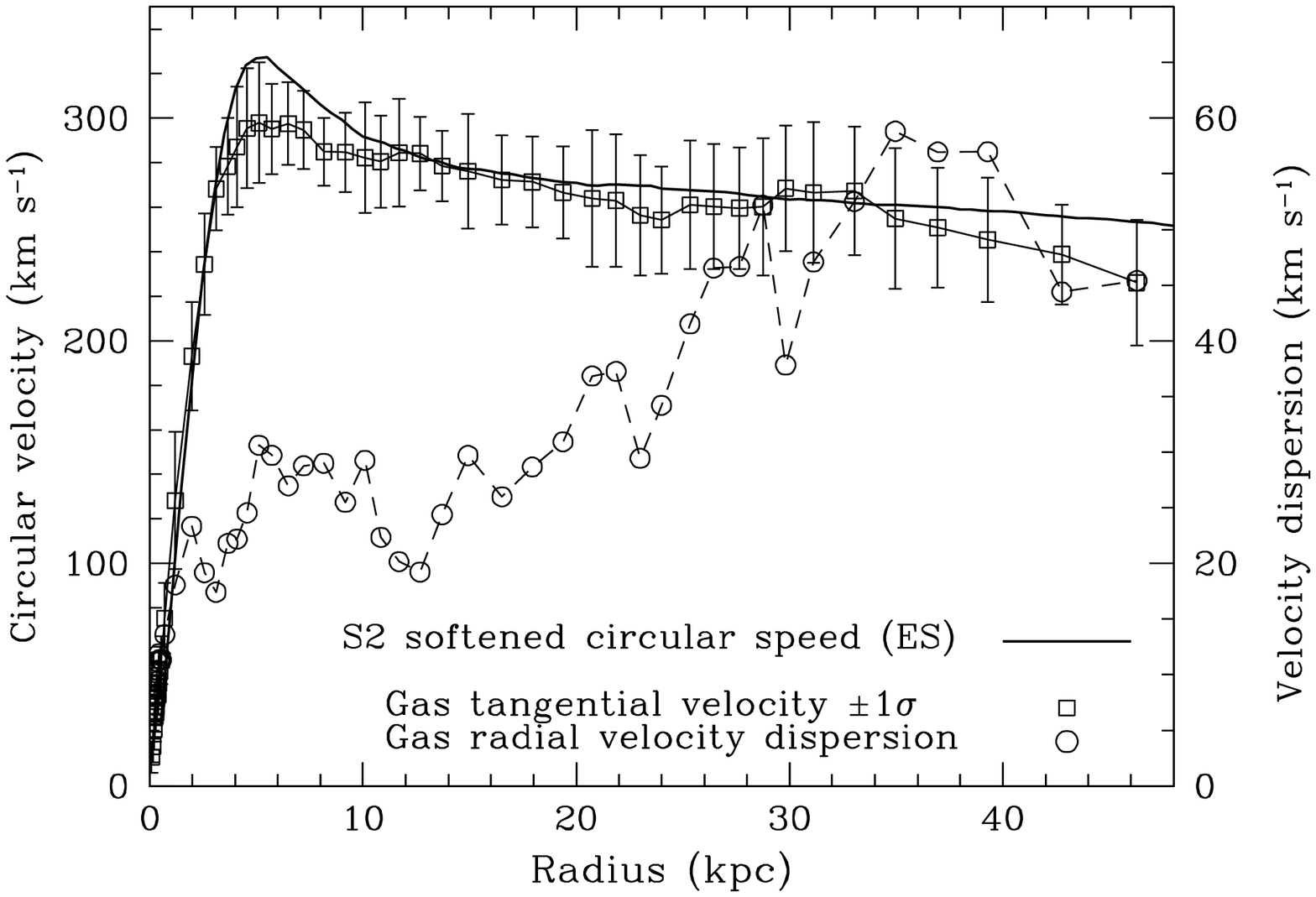}}

{\small {\sc Fig.}~3.---Tangential velocity and radial velocity dispersion
for the ES galaxy at z=0.52, averaged within 104 particle Lagrangian bins.
The tangential velocities compare well with the expected circular speed.}

\subsection{Rotation curve and density profiles}\label{rcden}
The large size of the stellar bulge makes an accurate determination of the
stellar disk rotation curve difficult. As a compromise we concentrate on
the gas disk rotation curve where star formation is ongoing. In
\fig~3 we show the tangential velocity of the gas in the ES run,
compared to the expected rotation curve, which has been softened using the
S2 softening shape (see TC00 for details). Rotation curves for the runs
are broadly similar, peaking at 327(312;340)\t\kms. The NF run has the
lowest central density as the rapid creation of stars during the formation
process results in a comparatively diffuse central core, which is an
unavoidable artifact of the `low' resolution of our models. Radial dark
matter density profiles are almost identical, with $\rho(r)\propto r^{-2}$
interior to $r=6$ kpc, similar to the results of Tissera \&
Dominguez-Tenreiro (1998). A run without baryons has a profile that
matches the Moore \etal (1999) fit. 

\section{Summary and Discussion}
Attempts to model star formation and feedback in a consistent way in
previous cosmological simulations of galaxy formation have had limited
success because the minimum simulation timescale far exceeds the
characteristic physical feedback time. This mismatch prevents feedback
energy coupling effectively to the system; energy which is therefore
unavailable to regulate star formation in the merging hierarchy.  We
have shown that by including a model for supernovae feedback in our
simulation which allows energy to persist for a timescale of 30 Myr,
we have overcome this obstacle. The technique has resulted in the
preservation of half of the specific AM content of the baryons in the
galaxy during collapse, 10\% lower than the value required to
reproduce observed disks at $z=0$. (Some part of this deficit may be 
viewed 
as
being due to the high $V_c/V_{200}$ ratio in sCDM.) The disk
scale length in the run with feedback is found to be within 17\% of
that predicted on theoretical grounds assuming the same fraction of
specific AM loss. The inclusion of star formation alone does not help
the AM problem for this halo. Further, without feedback, there is a
rapid exhaustion, by $z\sim0.5$, of the gas supply for star formation:
the overcooling problem in CDM cosmologies. Reducing the star
formation efficiency is not a solution as the specific AM values will
remain similar to the NSF run.

The effects of varying resolution are difficult to conjecture. The
assumption of the appropriate---but fixed---physical time-scales is a
barrier to the development of a resolution-independent model.  It is
worth noting, however, that our feedback prescription will always
produce feedback regions of a characteristic temperature between
$10^5$ and $10^6$ K (see TC00). At higher resolution the first
halos form with lower escape velocities and are
consequently more susceptible to the effects of feedback, perhaps
helping reduce the high central mass concentration evident
in \fig~3. Authoritative answers await higher resolution
simulations. We plan to conduct a simulation with twice the linear
resolution and eight times the mass resolution.  We are also
conducting a `survey' simulation using this feedback model at lower
resolution, comparable to Evrard \etal (1994).

The galaxy that we have simulated is large and represents a relatively
rare galactic event and it is unwise to extrapolate the
success of this particular model halo to a more general endorsement of
CDM cosmogonies. The key result of this paper, however, is that
hierarchical structure formation in not an anti-requisite for the
successful formation of discs: the adoption of a plausible physical
model for feedback can indeed regulate star formation and avoid
catastrophic angular momentum loss.

\acknowledgements
The authors thank Marc Davis for comments on the draft manuscript. A
grant of time on the UK-CCC server, `COSMOS', provided by the Virgo
Consortium is acknowledged.  This research was supported by NSF KDI
Grant 9872979 and NSERC (Canada). HMPC thanks the Canadian Institute
for Advanced Research for support.

\end{document}